\documentclass[
 aip,
 jcp,
 reprint,
 author-numerical,
]{revtex4-1}

\usepackage{graphicx} 
\usepackage{tablefootnote}
\usepackage{hyperref}
\usepackage{geometry}
\usepackage[T1]{fontenc}
\usepackage{amsmath}
\usepackage{multirow}
\usepackage{nicematrix}
\usepackage{xcolor}
\usepackage{float} 
\usepackage{notes2bib}

\begin{document}
\newcommand{\bo}{B$_2$O$_3$}


%
\title{Slowly Quenched, High Pressure Glassy B$_2$O$_3$ at DFT Accuracy}

\author{Debendra Meher}
\affiliation{Chemistry and Physics of Materials Unit, Jawaharlal Nehru Centre for Advanced Scientific Research, Bangalore 560064, India }

\author{Nikhil V. S. Avula}
\affiliation{Chemistry and Physics of Materials Unit, Jawaharlal Nehru Centre for Advanced Scientific Research, Bangalore 560064, India }

\author{Sundaram Balasubramanian}
\email{bala@jncasr.ac.in}
\affiliation{Chemistry and Physics of Materials Unit, Jawaharlal Nehru Centre for Advanced Scientific Research, Bangalore 560064, India }

\date{\today}

\begin{abstract}Modeling inorganic glasses requires an accurate representation of interatomic interactions, large system sizes to allow for intermediate-range structural order, and slow quenching rates to eliminate kinetically trapped structural motifs. Neither first principles- nor force field-based molecular dynamics (MD) simulations satisfy these three criteria unequivocally. Herein, we report the development of a machine learning potential (MLP) for a classic glass, \bo\, which meets these goals well. The MLP is trained on condensed phase configurations whose energies and forces on the atoms are obtained using periodic quantum density functional theory. Deep potential MD (DPMD) simulations based on this MLP accurately predict the equation of state and the densification of the glass with slower quenching from the melt. At ambient conditions, quenching rates larger than 10$^{11}$ K/s are shown to lead to artifacts in the structure. Pressure-dependent X-ray and neutron structure factors from the simulations compare excellently with experimental data. High-pressure simulations of the glass show varied coordination geometries of boron and oxygen, which concur with experimental observations.
\end{abstract}
\maketitle
\section{Introduction}


Boron trioxide (\bo) exists in one of three solid forms: two crystalline forms and the more common glassy \bo. Crystalline \bo-I consists of three-coordinated boron consisting of BO$_{3}$ units forming infinite chains~\cite{gurr1970crystal,burianek2016high}, while \bo-II features four-coordinated boron with corner-shared BO$_4$ tetrahedra stabilized at high pressures~\cite{prewitt1968crystal}.  
Glassy \bo\ under ambient conditions consists of a disordered network of planar BO$_3$ groups, with intermediate-range order through a six-membered planar boroxol rings~\cite{Brazhkin2010Viscosity,pasquerello_prl_2005}. The vitrification of \bo\ is rather facile. In contrast, the low-pressure crystalline phase, B$_2$O$_3$-I, has never been crystallized from its melt at ambient conditions, a phenomenon termed as the ''\bo\ crystallization anomaly''~\cite{aziz1985crystal,ferlat2012hidden,ferlat2019van}. Over the years, glassy \bo\ has been extensively researched due to its unique properties, including low thermal expansion, weak electrical conductivity, high resistance to thermal shock, and strong corrosion resistance~\cite{fundamentlsglass}.

A major impediment to understanding the properties of glasses is the lack of access to their structures at the atomic scale. Often, many structural models appear to fit available experimental data reasonably (one or more of the physical properties, scattering functions, NMR chemical shifts, EXAFS, etc.), thus making the selection of a particular model as the ground truth difficult. However, atomistic simulations based on interatomic interactions can and do provide microscopic structural details. While empirical force fields have been developed for many inorganic glasses, they are often parameterized to reproduce specific properties and may not accurately capture the true physics of the system. In contrast, ab initio molecular dynamics (AIMD) based on first principles (primarily quantum density functional theory (DFT)) can produce outcomes consistent with experimental data. Still, they are limited by relatively small system sizes (hundreds of atoms) and short trajectories. The former curtails the observation of possible intermediate-range order in glassy networks. The latter impedes the exhaustive sampling of configurational space even in the melt, a necessity for atomistic simulations to be considered as being in thermodynamic equilibrium.
Notwithstanding these intrinsic limitations, in the specific case of glassy B$_2$O$_3$, AIMD simulations have successfully captured the coordination transformation of boron from three- to four-coordinated states under increasing pressure~\cite{JPCC_high_press_2024}. Further, they have also been used to characterize the \bo\ melt, and derive three-body potentials to study the glass structure and dynamics~\cite{SCHERER201973}. 
However, AIMD is computationally expensive and employs extremely high quenching rates, at least 8-10 orders of magnitude larger than routinely used in experiments.

To address these challenges, Machine Learning Potentials (MLPs) based on DFT data have been developed, enabling the prediction of structures and dynamics with a speed increase of up to a factor of one thousand. The best practices for the development and reporting of such MLPs have been discussed in the literature~\cite{morrow2023validate,tokita2023train}. Deep Potential Molecular Dynamics (DPMD) simulations based on MLPs have been employed to understand the structure and thermodynamics of various inorganic materials, such as boron phosphide at high pressure~\cite{Boron_phosphide_2023}, bulk TiO$_2$ at high pressures~\cite{liu2024deep}, amorphous alumina~\cite{aAlumina2020}, metal oxides~\cite{Sivaraman2022}, thermal conductivity in silica~\cite{SilicaPegolo2024} and the melting of boron nanoparticles~\cite{Dongpingchen2023}.  The literature on MLPs to model liquid water and aqueous solutions is vast, and a recent perspective provides a succinct summary~\cite{omranpour2024perspective}.

Although  MLPs\cite{Bertani2024,Urata_LBS_2022,Urata_silica_2022,Urata_LBG_2024,Dongpingchen2023,ChemPhysChem2024} have been constructed for borosilicate glasses, that for neat \bo\ glass has not been explored yet. In the present work, we report the development of an MLP for pure glassy \bo\ trained on DFT data, which enables its study at ambient and high-pressure conditions. Our findings indicate that the high quenching rates typically used in AIMD simulations can potentially introduce artificial structural motifs not present in samples quenched conservatively. Further, we report the increased densification of glassy \bo\ with a decrease in quenching rate, an observation that is consistent with experiments~\cite{SOPPE1988201}, which reported \bo\ glasses of densities 1.79 g/cc and 1.83 g/cc at quenching rates of 10$^{3}$ K/s and 4$\times$10$^{-3}$ K/s respectively. The DPMD simulations also reveal the transformation of BO$_3$ to BO$_4$ units beyond a glass density of 2.4 g/cc (corresponding to around 5.5 GPa). The structural models reproduce all the features of the experimentally reported neutron and X-ray structure factors and their dependence on pressure.

\maketitle
\section{Simulation Details}

\subsection{Machine Learning Potential for \bo}
In this work, machine learning potentials that represent the potential energy surface of the liquid and glassy phases of \bo\ at the accuracy of quantum density functional theory (DFT) have been developed. Specifically, a generalized gradient approximation (GGA) to the DFT has been adopted, which has earlier been used to \textit{reliably} describe many properties of oxide glasses \cite{waltorkob_glass} including \bo \cite{Brashkin_2008_AIMD}. Further, a dispersion correction (Grimme's D3) to the revPBE functional was also added to overcome the general drawback of GGA functionals underestimating the mass density of materials. Ferlat et al. observed that adding dispersion corrections to GGA functionals makes their predictions of \bo\ crystal density comparable to that of high-level quantum calculations like quantum Monte Carlo (QMC), giving further credence to our choice of theory \cite{ferlat2019van}. The same point has been underscored by Assaf et al~\cite{assaf2017structure}.

This section describes the dataset generation, training, and validation of the MLPs.

\subsubsection{Generation of the Training Dataset}
The training dataset of MLPs should typically include frames sampled from the thermodynamic phases and state points of interest.  The main focus of this work is the glassy B$_2$O$_3$ phase at room temperature and high pressures up to 20 GPa. Hence the training set contains frames sampled from the liquid and glass phases with temperatures ranging from 5 to 3000 K and pressures ranging from 0 to 500 GPa. 

Active learning procedures are generally used to create the training dataset for MLPs in an efficient manner \cite{Zhang2019PRM,Vandermause2020,DPGEN,Boron_phosphide_2023}. In this procedure, a few exploratory MD runs are carried out with an ensemble of MLPs, and configurations are selected based on the uncertainty of the MLP ensemble. However, in the case of the \bo\ system, the dynamics are so sluggish that the configurations from an exploratory run tend to be highly correlated. To overcome this, the configurations were sampled from many independent exploratory runs each starting from different randomly packed structures.

Overall, the training dataset was generated in two stages. In the first stage, the configurations were sampled from exploratory MD  runs with a classical empirical potential \cite{WANG2018} at elevated temperatures (1600 to 3000 K). This process was repeated for six generations until a stable MLP was obtained at 1600 K and 1.502 g/cc (corresponding to the experimental liquid density). These MLPs were used in the second stage to sample configurations from other state points in an iterative manner, with each successive generation of MLPs covering a wider range of state points. After getting a stable MLP across state points (MLP generation \#20), the force field-based MD simulation frames were discarded. The contributions of frames at different state points to the training dataset are listed in Table S2. 
\begin{figure}[ht]
     \centering
     \includegraphics[width=1\linewidth]{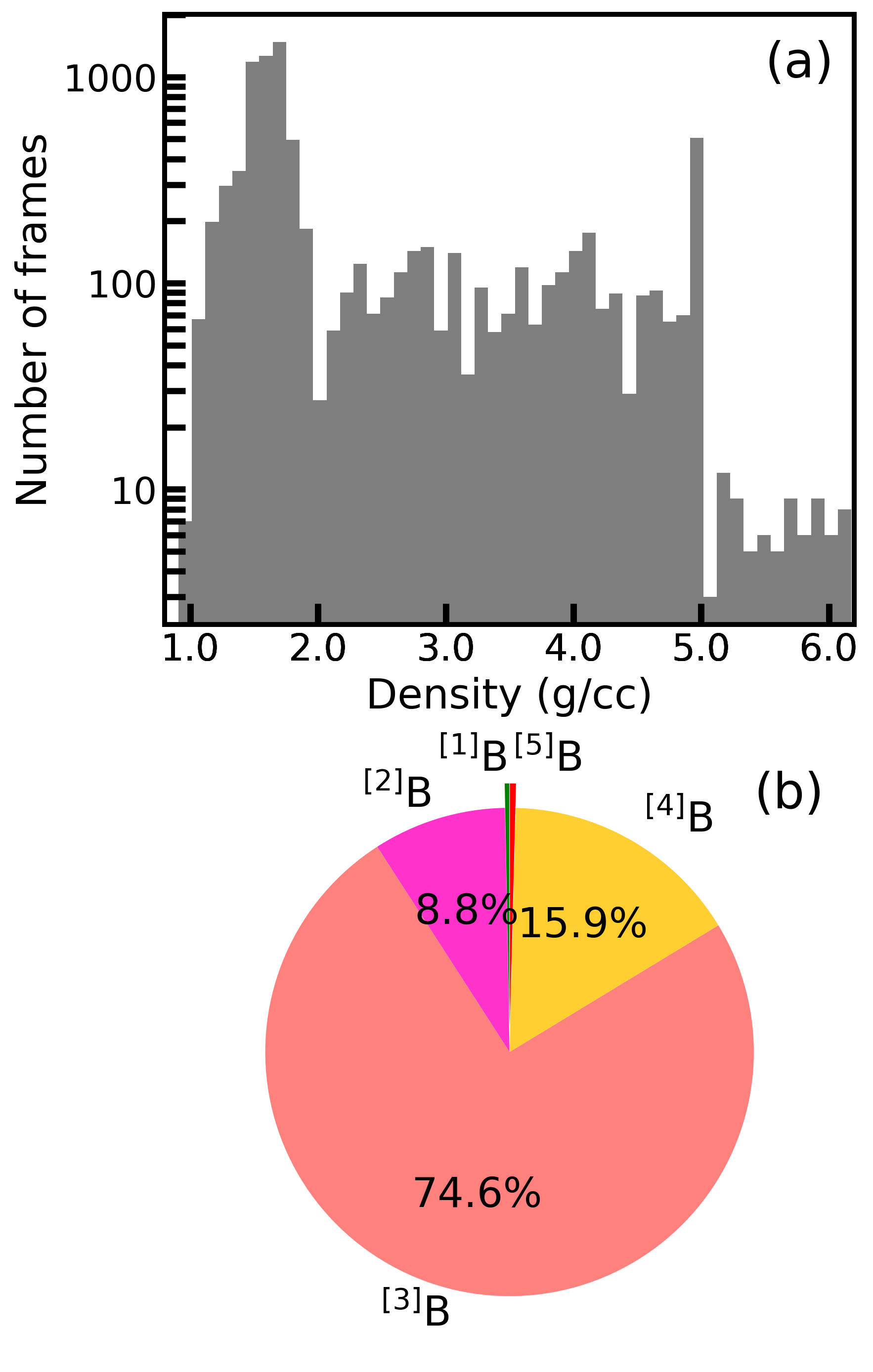}
     \caption{(a) Distribution of the mass density of the configurations sampled for the training set. (b) The pie chart displays the distribution of boron atoms in the training set with different (one to five) coordination numbers. In $^{[n]}$B, $n$ is the number of oxygen atoms coordinated to boron. The experimentally reported density of \bo\ glass at ambient conditions prepared with a quenching rate of 10$^{-3}$ K/s is 1.83 g/cc.~\cite{SOPPE1988201}}
     \label{fig:density_hist}
\end{figure}   

Figure \ref{fig:density_hist}(a) shows the distribution of the mass density of the configurations in the final dataset highlighting the wide range of values from 1.0 to 6.0 g/cc.  As indicated earlier, boron atoms can change coordination (both in terms of number and geometry) depending on the thermodynamic conditions \cite{PRL_Braz_EXP_2008_high_dens,IXS_nature,Salmon_2015,PNAS_x_ray_2018,high_pressure_review_APL_2022,JPCC_high_press_2024} and hence the training dataset must contain different coordination environments. Figure \ref{fig:density_hist}(b) shows the distribution of boron coordination numbers in the final dataset. This indicates a significant presence of two-, three-, and four-coordinated boron atoms and a few one- and five-coordinated ones. The final dataset contains around 9500 configurations or equivalently 9500 energy and 7552500 force component values, respectively. The dataset was shuffled and split into training (90\%) and validation (10\%) sets. Each configuration in the dataset contains 53 units of \bo\ or 265 atoms. 

The energy, forces, and virial of a configuration were computed using the Quickstep module of the CP2K program \cite{cp2k}. The calculations employed the revised version of the Perdew-Burke-Ernzerhof (revPBE)\cite{revPBE} exchange-correlation functional, and the valence electrons were described with a molecularly optimized Double-Zeta Valence and Polarization with Shorter-Range (DZVP-MOLOPT-SR)~\cite{molopt_basis} basis set. The core electrons and the nuclei were treated using the Geodecker-Teter-Hutter (GTH) pseudopotential \cite{GTH1, GTH2}. Grimme's D3 empirical corrections \cite{dispersion-D3}, with a cutoff of 40 \AA\, were applied to include dispersion contributions. 

A high electron density cutoff of 1200 Ry was used to obtain a well-converged energy (see SI Figure S3). The self-consistent field (SCF) iterations were considered converged if the energy difference between successive iterations was less than 10$^{-7}$ Hartree. Periodic boundary conditions were applied for all single-point energy calculations. Further details about the dataset generation are given in the supplementary material Section S1.

\begin{figure*}[ht]
    \includegraphics[trim=2cm 0cm 2cm 1cm, clip, width=1\textwidth]{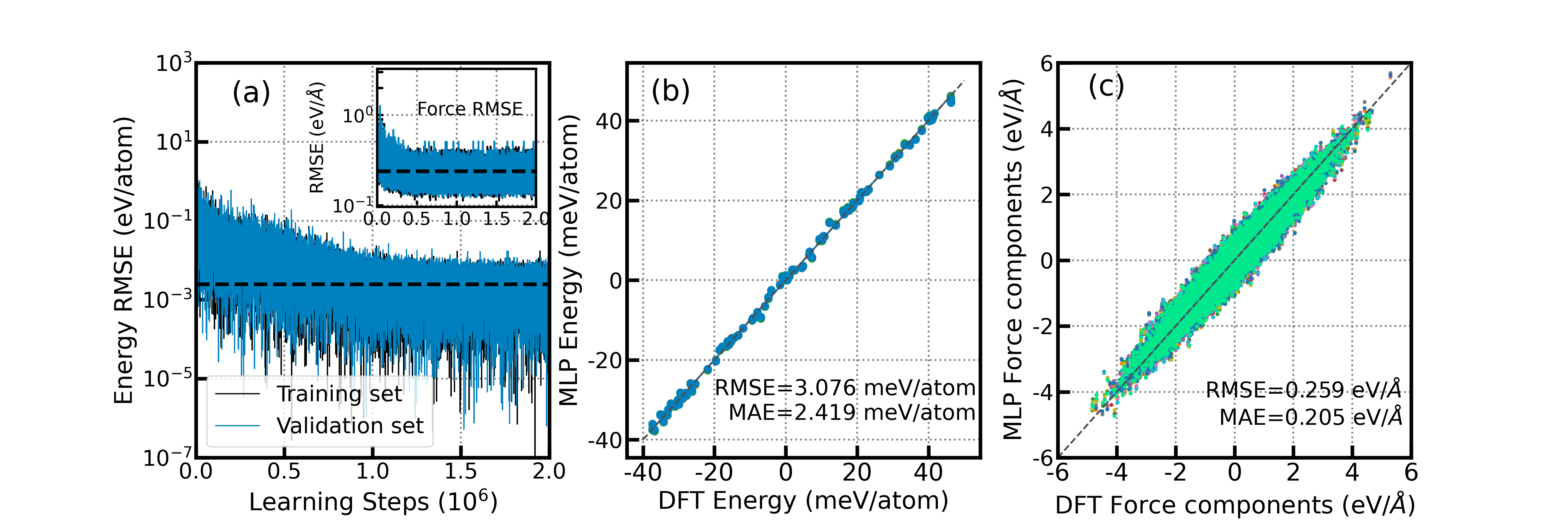}
    \caption{ (a)  RMSE energy plot for training and validation. Inset: the same for force. The blue dashed line is the average value of the last 5\% of the validation set. (b) Parity plot of the potential energy of the configurations from four different independent MLPs, where DFT energy is along the abscissa, and the predicted energy by the MLP is along the ordinate. (c) Force components (F$_x$, F$_y$, F$_z$) are plotted for the four independent MLP models, each colored differently.}
    \label{fig:MLP_performance}
\end{figure*} 

\subsubsection{MLP Details}
In this work, we utilize the Deep Potential (DP) framework, based on an end-to-end neural network architecture, to create a machine learning potential for \bo\ system. To create the DP model, we employed the DeePMD-kit\cite{deepmd,deepmdv2} software, which utilizes the Deep Potential Smooth Edition (DeepPot-SE) construction. This model incorporates embedding and fitting networks powered by separate neural networks. The embedding network learns the local environment within a cutoff radius around an atom, while the fitting networks learn the total energy of the system and are fed the local atom embeddings obtained from the embedding network. For more details about the DP architecture and its implementation, readers are referred to a recent article from its developers \cite{deepmdv2}.

The embedding network was configured with a size of [25,50,100] and a cutoff of 6 Å. The fitting network consisted of three hidden layers, each containing 250 nodes. Table S3 shows the variation of training and validation errors with other DP hyperparameters - $r_s$, number of training steps. To train the network, we utilized two million steps with a learning rate exponentially decreasing from 5 $\times$ 10$^{-3}$ to 1.76 $\times$ 10$^{-7}$ and a decay rate of 5000 steps. The loss function was computed using a weighted sum of energy and force errors, with the energy and force prefactors that varied throughout training. Figure~\ref{fig:MLP_performance}a shows the variation of energy loss (root mean squared error (RMSE)) on the training and validation sets over the training steps. The training and validation errors are very close, indicating the absence of over/underfitting. Also, similar neural network architectures have been used by DP models to study aqueous electrolytes \cite{zhang2022dissolving,avula2023understanding}, 
molten salts \cite{lam_molten_salt2022}, inorganic glasses \cite{Boron_phosphide_2023,Urata_LBG_2024}, etc. Finally, four different MLPs were trained with the same model architecture starting from different initial weights to enable the estimation of model deviations over the MD trajectories (see below).

\begin{figure*} 
    \centering
    \includegraphics[trim=2cm 0cm 2cm 0cm, clip, width=1\linewidth]{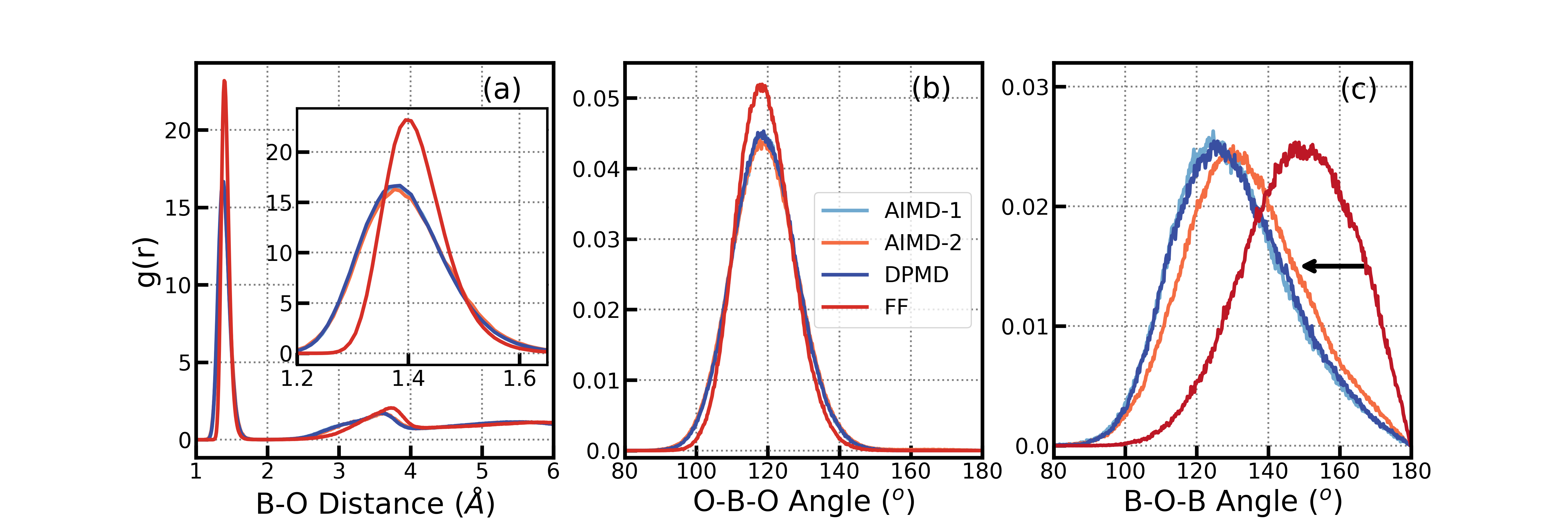}
    \caption{Comparison of DPMD results against those from two AIMD simulations started from different initial configurations, all performed at 1.49 g/cc and 2000 K. Results from empirical force field-based MD simulations are also compared. (a) B-O RDF (b) O-B-O angle distribution (c) B-O-B angle distribution. The AIMD simulation, whose result is shown in orange (AIMD-2), was started from a configuration equilibrated with the force field (FF), whose distribution is in red; the arrow represents the time evolution of the distribution. On the other hand, the AIMD-1 simulation was started from a configuration equilibrated by DPMD. In panel (c), the orange curve does not overlap with the results of either DPMD or AIMD-1 due to the short duration of the trajectory (40 ps).}
    \label{fig:AIMD_comparison}
\end{figure*}

\subsubsection{Accuracy of the MLPs}
Figure \ref{fig:MLP_performance}b,c shows the final MLP (generation \#26) predicted energy and force parity plots evaluated on the validation dataset. The training and validation root mean square error (RMSE) of energy and force were 2.5 meV/atom, 240 meV/\AA, and 2.6 meV/atom, 247 meV/\AA, respectively. The accuracy of our MLPs compares well with those reported for other MLPs of oxide glasses \cite{cooling_rateJCP2019,aAlumina2020}. Recently, other complementary metrics were proposed to better evaluate MLPs based on structure and dynamics comparison with the corresponding AIMD simulations \cite{fu2023forces,EGraFFBench}. In that vein, to validate the DPMD trajectory, we carried out two AIMD simulations of the \bo\ melt and compared the results of the structure between them. 

{\bf Validation against AIMD:}  Due to the computational challenges of AIMD simulations and the need for adequate sampling, we conducted these simulations at 2000 K for a small system size of 125 atoms (25 \bo\ units). The AIMD simulations were started from two different initial configurations, as they were short in duration. The system's density was 1.49 g/cc, and constant NVT conditions were adopted. The first AIMD simulation (AIMD-1), lasting 20 ps, began from an initial configuration well-equilibrated with DPMD at 2000 K. The second AIMD simulation (AIMD-2), lasting 40 ps, started from a configuration pre-equilibrated using the empirical force field of Wang et al.\cite{WANG2018}. Figure~\ref{fig:AIMD_comparison} compares the radial distribution function (RDF) of B-O, as well as the angle distributions of B-O-B and O-B-O, across the four MD simulations. These include the AIMD and DPMD simulations, alongside the classical MD simulation, all initiated from their respective equilibrated configurations. The structural properties predicted by AIMD and DPMD are in excellent agreement, even when starting from configurations equilibrated with the classical force field. In contrast, deviations observed in the RDF and bond angle distribution for the classical MD simulation indicate that the classical force field is less accurate in modeling the melt (and, thus, expectedly, the glassy \bo\ system).

{\bf Stability of DPMD:} We tested the final DP model (generation \#26)  across different temperature and pressure ranges (0-3000 K and 0-500 GPa) and found the DPMD trajectories stable over 10 ns. The stability of the MD trajectories was monitored using structural analysis (close atom contacts etc.), and MLP ensemble uncertainty ($\zeta$) was estimated using Eq. \ref{Eq:model_devi1}. Figure S2 shows the MLP ensemble uncertainty of a DPMD run of \bo\ system containing 1700 atoms (340 \bo\ units) at 2400 K and 3.6 g/cc density over 10 ns. In the current manuscript, we limit the analysis of the results to 300 K temperature and pressures up to around 20 GPa.

\begin{center}
\begin{equation}
    \zeta = max \sqrt{\langle | \textbf{F}_{i} - \langle \textbf{F}_{i} \rangle |^{2} \rangle} 
\label{Eq:model_devi1}
\end{equation}
\end{center}
Here, $\textbf{F}_{i}$ is the force on atom i, and $\langle \rangle$ represents the average over the MLP ensemble. 

\subsection{Deep Potential Molecular Dynamics Details}
We ran all the Deep Potential Molecular Dynamics (DPMD) simulations using the LAMMPS package \cite{lammps} patched with DeePMD-kit \cite{deepmdv2}. All DPMD simulations were monitored for stability and correctness using the model uncertainty $\zeta$ (Equation~\ref{Eq:model_devi1}).

We utilized PACKMOL\cite{packmol1,packmol2} to randomly pack 1700 atoms, equivalent to 340 \bo\
units at the target density. Subsequently, we conducted energy minimization using the empirical force field~\cite{WANG2018} to achieve a more stable arrangement.
\bibnote{The random configuration generated by PACKMOL contained, occasionally, unphysical hard contacts that could not be relaxed with the MLP, and thus, the force field was used to perform energy minimization.}
Following this, we began a DPMD simulation under constant NVT conditions for
each density, starting from 2400 K. The 
van Hove correlation function (VHCF) (Figure
S6)  and mean square displacement (MSD) (Figure S7) 
indicated that at 2400 K, atoms moved more than 10 \AA\ in 1 ns
and are well mixed in the MLP description, which is ideal for configurational
sampling. After a 5 ns equilibration at 2400 K, we gradually decreased the
temperature to 300 K at different quenching rates (10$^{11}$ K/s, 10$^{12}$
K/s, 10$^{13}$ K/s, 10$^{14}$ K/s), and then equilibrated for another 1 ns at
300 K. A final 1 ns production run was performed for all the analyses. Longer
production runs were not necessary due to the viscous nature of the system,
which limits configurational sampling at 300 K. The above procedure was carried
out for four initial configurations at each density, amounting to four
independent glass networks at 300 K per density studied.

The history of system preparation is crucial in determining the final simulation results for glassy materials. Factors such as quenching rate, high temperature, and pressure during quenching can significantly impact the outcomes. An earlier simulation study on amorphous silicon explored quenching rates ranging from the slowest at 10$^{11}$ K/s to faster rates at 10$^{12}$ K/s, 10$^{13}$ K/s, and 10$^{14}$ K/s\cite{Quenchrate_Si}. It revealed that the slowest quenching rate produced structural properties closest to experimental results. Another study by Li et al. utilized machine learning potentials to generate different amorphous silicon structures by controlling the quenching rate and matched these structural properties to different experimental observations \cite{cooling_rateJCP2019}. 
Herein, we conducted a similar investigation by employing different quenching rates under constant volume conditions to obtain glassy \bo\ of various densities at 300 K. Subsequently, the dependence of the glass structure on the quenching rate was studied.

To our knowledge, we have followed the best practices of developing MLPs as advocated in the literature~\cite {morrow2023validate,tokita2023train}. The MD trajectories were analyzed using VMD~\cite{vmd} and home-grown codes. The structure factor was calculated using  TRAVIS~\cite{travis}, while the ring statistics in the glassy network were obtained using networkx~\cite{networkx}. A distance cutoff of 1.7 \AA\ was used to determine the B-O bonded pairs, which are, in turn, used for coordination environment and network analyses.

\begin{figure}[H]
    \centering
    \includegraphics[width=0.82\linewidth]{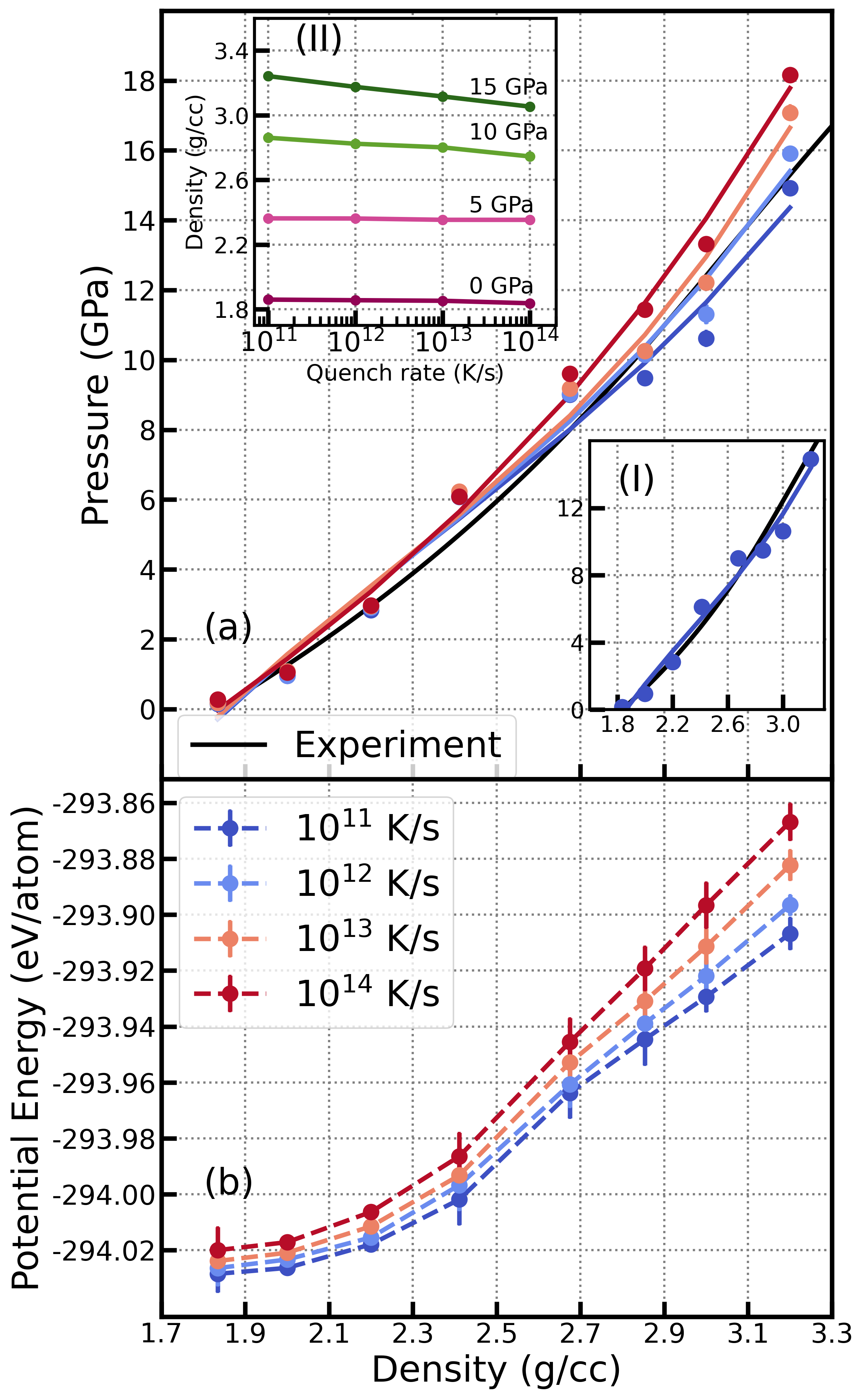}
    \caption{(a) Pressure as a function of density at 300 K for glassy \bo\ obtained through DPMD at four different quenching rates is compared with experimental\cite{PRL_Braz_EXP_2008_high_dens} data. The solid lines are third-order polynomial fits. The error bars are not visible as the errors are smaller than the symbol size. Inset I: Comparison of P-$\rho$ for the glass quenched at 10$^{11}$ K/s with experimental data~\cite{PRL_Braz_EXP_2008_high_dens}. Inset II: Dependence of \bo\ glass density on quenching rate for various pressures. (b) Potential energy per atom as a function of glass density at different quenching rates.  The dashed lines are a guide to the eye.}
    \label{fig:density_vs_pressure}
\end{figure}
\section{Results and Discussion}
\subsection{Quenching Rates} 
\subsubsection{Density and Potential Energy}

\begin{figure*}
    \centering
    \includegraphics[trim= 0cm 0cm 0cm 0cm, clip,width=1\linewidth]{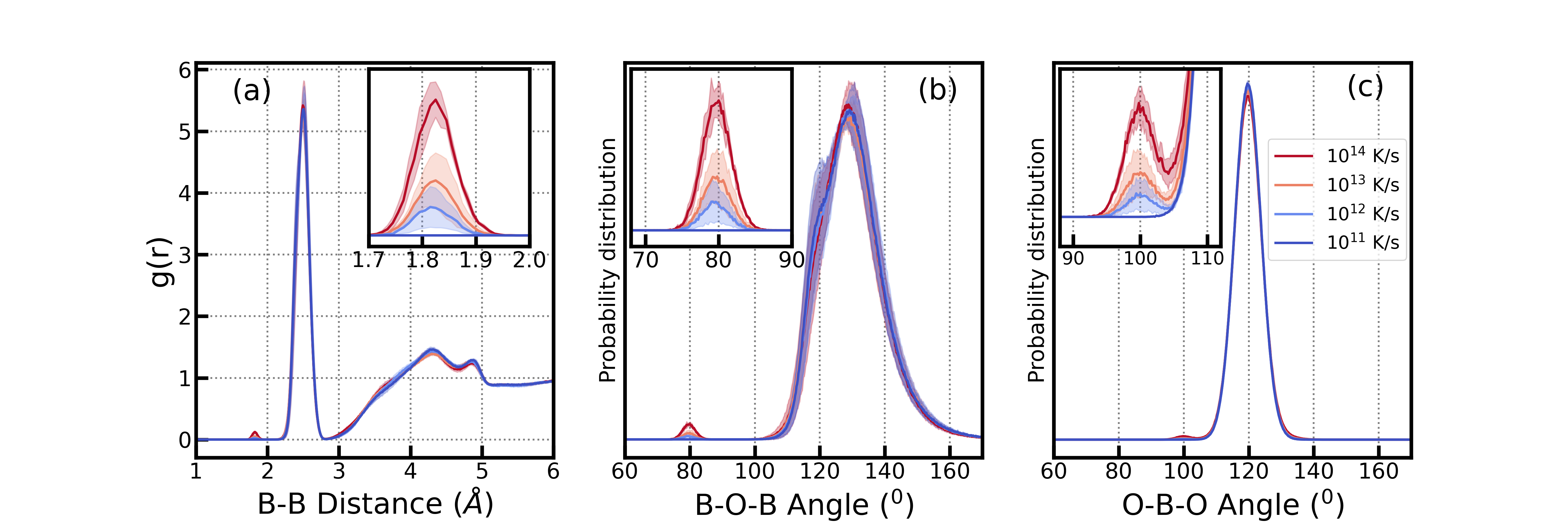}
    \caption{Effect of quenching rate on the structure of the glass at 300 K and 1.834 g/cc density. (a) B-B RDF shows a small peak at a distance less than 2 \AA\ at high quenching rates. Its height decreases with a decreasing quenching rate, and the peak vanishes at the slowest quenching rate of 10$^{11}$ K/s. (b) B-O-B bond angle distribution displays a hump at 80$^\circ$ for high quenching rates, which is the angle formed at the oxygen atom in a four-membered ring (B-O-B-O) with a boron-boron distance of around 1.8 \AA. (c) the bond angle distribution of O-B-O too shows a peak at $100^\circ$ due to the presence of a few four-membered rings formed as an artifact at high quenching rates; the same is absent at the slowest quenching rate. Insets in all the panels show the zoomed-in region around this artifact. The shaded region around lines shows the standard deviation over four independent trajectories.}
    \label{fig:quench_rate_comparison}
\end{figure*}

The density of glassy materials significantly influences their structure and other properties. Achieving a well-relaxed system typically requires a lengthy equilibration run. Our simulations of B$_2$O$_3$ glass at 300 K were conducted under constant NVT conditions and were quenched from the melt at constant density conditions. The glasses at 300 K can thus be considered ''pressure-quenched''~\cite{smedskjaer2014irreversibility}. The corresponding pressure of the glass at 300 K reported by DPMD simulation, averaged over the four independent runs at each density, is presented in Figure~\ref{fig:density_vs_pressure}a. 
At low density, the pressure does not seem to depend on the quenching rate. However, with the increase in density, beyond say, 2.1 g/cc, the pressure depends on the same. The pressure obtained with the slowest quenching rate (10$^{11}$ K/s) is closer to the experimentally determined equation of state of the glass (Inset I of Figure~\ref{fig:density_vs_pressure}(a)). Inset II displays the evolution of the mass density of \bo\ glass at different pressures with quenching rate. At 0 GPa, the dependence is weak, while it is strong at higher pressures; decreasing quenching rate leads to denser glasses. In the case of the experimentally synthesized borate glass, the density at 298 K depends on the quenching rate -- varying from 1.79 g/cc at 10$^{3}$ K/s to 1.83 g/cc at 10$^{-3}$ K/s~\cite{SOPPE1988201}. Our results are consistent with such observations from experiments.

Figure~\ref{fig:density_vs_pressure}b displays the dependence of the average potential energy per atom on the glass density for various quenching rates. As anticipated, configurations with slower quenching rates display lower potential energy than those with faster quenching rates. This pattern persists across all densities. Nonetheless, at low densities, there is an overlap between the error bars, whereas at high densities, the distinction is clearer. Similar observations have been made by Deringer et al. in their study of amorphous silicon, wherein the slowest quenching rate of 10$^{11}$ K/s resulted in more stable (lower energy) structures than those obtained from quenching rates faster than 10$^{12}$ K/s \cite{Quenchrate_Si}.

\subsubsection{RDF and Bond Angle Distributions}
The radial distribution function (RDF) for atom pairs was calculated for four quenching rates and four different densities of \bo. The RDF of B-O showed consistent overlap across all quenching rates and densities, indicating that the local structure between boron and oxygen atoms remains largely unaffected by the quench rate or density variations. However, some discrepancies were observed in the RDF of B-B pairs.
Specifically, a small peak around 1.8 \AA\ was detected at a density of 1.834  g/cc, whose height diminishes as the quenching rate decreases. Figure~\ref{fig:quench_rate_comparison}(a) shows the RDF of B-B, with the inset highlighting a zoomed-in view of the small peak near 1.8 \AA. The literature has discussed this feature as a motif that arises in glasses formed with high quenching rates~\cite{WalterKob}.
Our analysis also found that this characteristic distance appears primarily in high quenching rate \bo\ samples at lower densities. This peak corresponds to the boron-boron separation within four-membered rings (B-O-B-O), artificially stabilized under rapid cooling conditions. As the quenching rate decreases, these four-membered rings become rare, reducing the intensity of the 1.8 \AA\ peak.

The signature of the four-membered rings can be observed in the bond angle distribution of B-O-B and O-B-O, too, as shown in Figure~\ref{fig:quench_rate_comparison}(b) and (c), respectively. These rings are characterized by a B-B distance of 1.8 \AA, a B-O-B angle of $80^\circ$, and an O-B-O angle of $100^\circ$. The boron atoms are three-coordinated at this density, and the oxygen atoms are two-coordinated. The small peak near 2 \AA\ diminishes in height as the density increases, even at the highest quenching rate (10$^{14}$  K/s). The change in coordination number can explain this behavior: at higher densities, boron, and oxygen coordination numbers increase from three and two to four and three, respectively. With this increase, the B-O-B bond angles shift from $80^\circ$ to $90^\circ$, and the B-B distance extends to 2 \AA. This angle at high density is evident in Figure~\ref{fig:ang_dist_press}(a).

Given that the glass structure at 1.834 g/cc density at quenching rates greater than and equal to 10$^{12}$ K/s contains an artifact, in the rest of the discussion, we present results of analyses of the 300 K glass obtained with a quenching rate of 10$^{11}$ K/s alone. Results from other quenching rates are discarded. 

\subsection{B$_2$O$_3$ Under Pressure}  
\subsubsection{Structure Factor}
\begin{figure}
    \centering
    \includegraphics[width=0.8\linewidth]{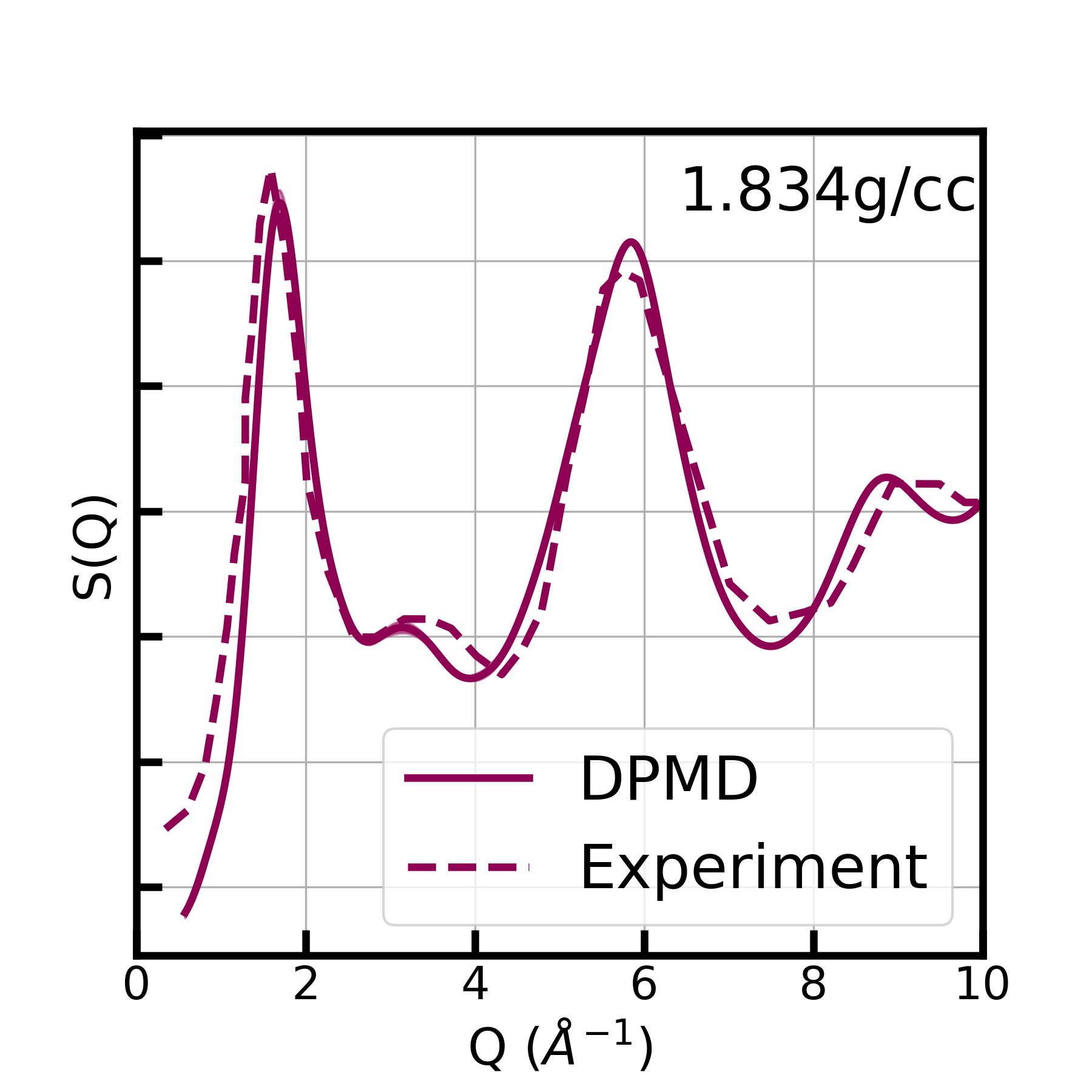}
    \caption{Comparison of the neutron structure factor of \bo\ glass at 300 K and 1.834 g/cc with the experimental data~\cite{DensityDrivenStruct2014}.}.
    \label{fig:neutron_structure}
\end{figure}
\begin{figure}[H]
    \centering
    \includegraphics[trim=1cm 2cm 1cm 1cm, clip, width=0.8\linewidth]{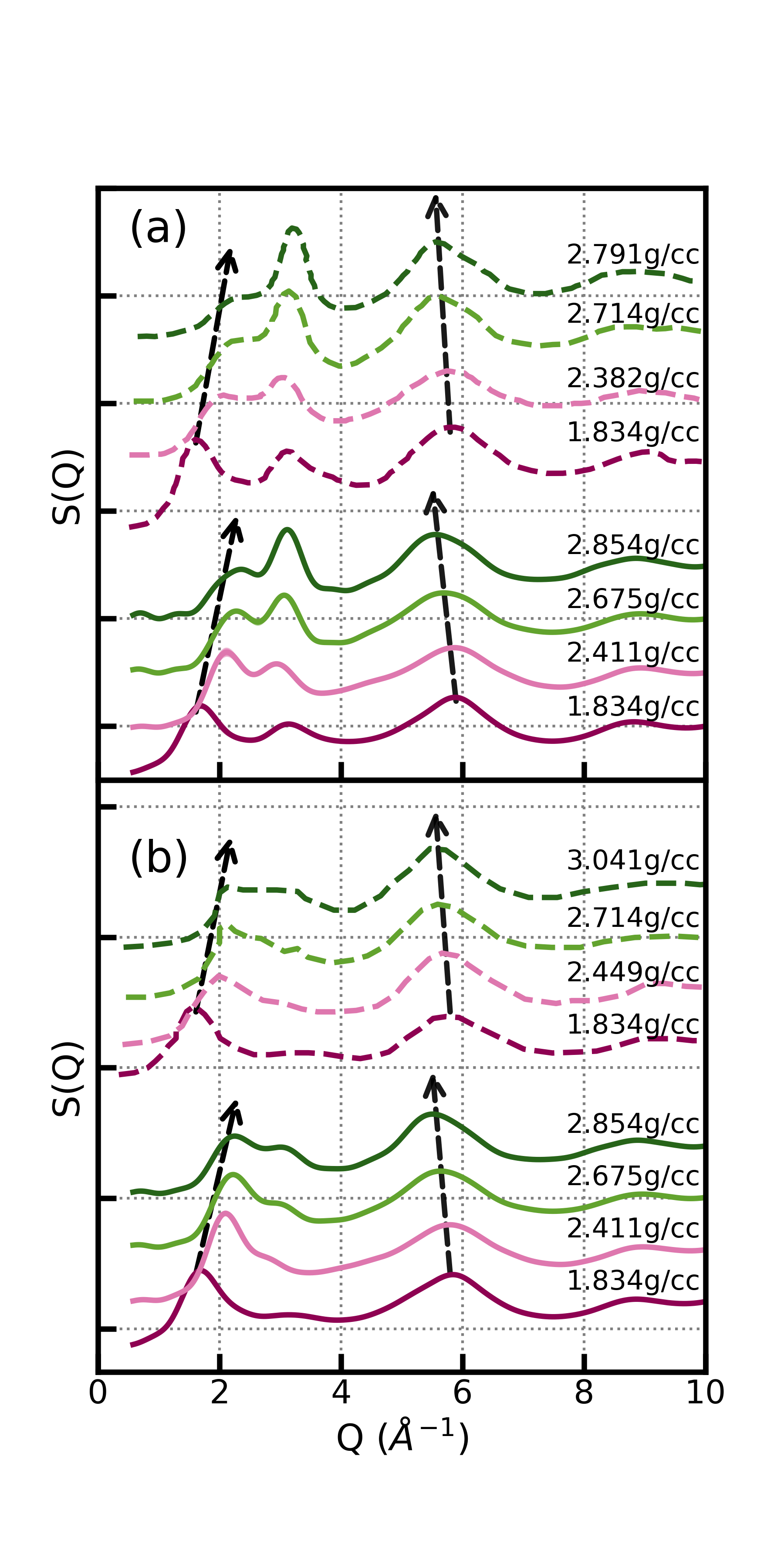}
	\caption{(a) X-ray structure factor of \bo\ glass at different densities. Dashed lines: Experiment~\cite{PRL_Braz_EXP_2008_high_dens}, Solid lines: DPMD simulations. (b) Neutron structure factor for different densities. Dashed lines: Experiment~\cite{DensityDrivenStruct2014}, Solid lines: DPMD simulations. The black dashed arrows are drawn as a guide to the eye. The standard deviation over four independent trajectories is not visible as it lies within the linewidth of the curves. }
    \label{fig:x_ray_struct}
\end{figure}

At 1.834 g/cc, the neutron structure factor of the simulation matches the experimental result~\cite{DensityDrivenStruct2014} quite impressively (Figure~\ref{fig:neutron_structure}). 
Figure~\ref {fig:x_ray_struct}(a) shows the X-ray structure factor at different densities; the results from DPMD align well with experimental observations~\cite{PRL_Braz_EXP_2008_high_dens}. At a density of 1.834 g/cc, the first peak in the X-ray structure factor is taller than the second in both the experimental and calculated results. As density increases, the first peak height decreases, and its position shifts to higher wave vectors, while the second peak grows in intensity, but its position remains stationary. Additionally, the third peak shifts to the left as density increases. These behaviors observed in the DPMD calculations closely match the experimental structure factor trends.

	\begin{figure*}[ht]
    \centering
    \includegraphics[trim=4cm 0cm -4cm 0cm, clip,width=1.2\linewidth]{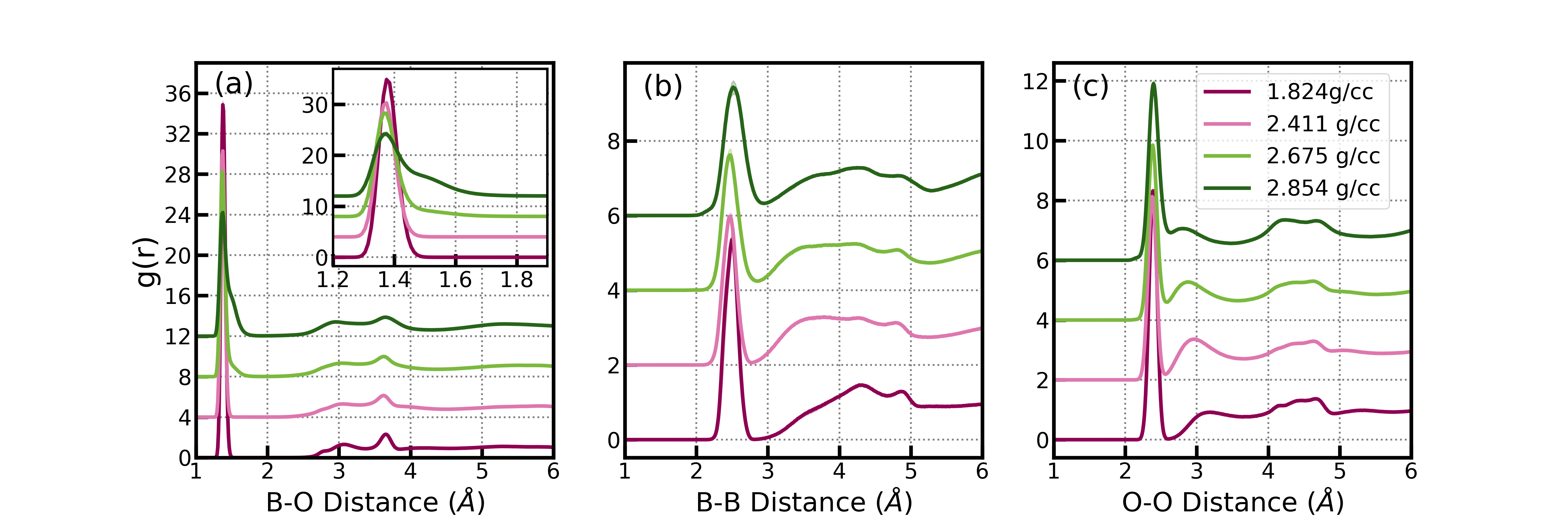}
    \caption{Comparison of RDF of \bo\ glass at 300 K at four different densities formed with the slowest quenching rate (10$^{11}$ K/s). (a) B-O (b) B-B (c) O-O. The standard deviation over four independent trajectories is not visible as it lies within the linewidth of the curves.
	The inset in panel (a) zooms into the region around the first peak.}
    \label{fig:rdf_dist_press}
\end{figure*}

Similarly, the neutron structure factor too was calculated for different densities and compared with the experimental data for the closest densities available~\cite{DensityDrivenStruct2014} in Figure~\ref{fig:x_ray_struct}(b). With increasing density, the second small peak merges with the first peak initially as a shoulder and later to broaden it at higher densities. Again, the third peak at 5.73 \AA$^{-1}$ at 1.834 g/cc shifts towards the left with increasing density, which mirrors the trend in the experimental neutron structure factor. 

\subsubsection{Radial Distribution Functions}
Figure~\ref{fig:rdf_dist_press} shows the RDFs for B-B, B-O, and O-O pairs at various densities. Experimental data~\cite{ExpB2O3struct} suggests the B-O bond length to be between 1.37 \AA\ and 1.40 \AA\ in \bo\ glass at ambient conditions. The current DPMD simulations yield a mean B-O bond length of 1.375 \AA. As the sample density increases, the height of this first peak decreases, and a shoulder emerges to its right. This shoulder is linked to the formation of four-coordinated boron atoms, indicating a change in the coordination environment of boron from planar triangular to tetrahedral geometry. 
	\begin{figure}[h]
    \centering
   \includegraphics[width=0.9\linewidth]{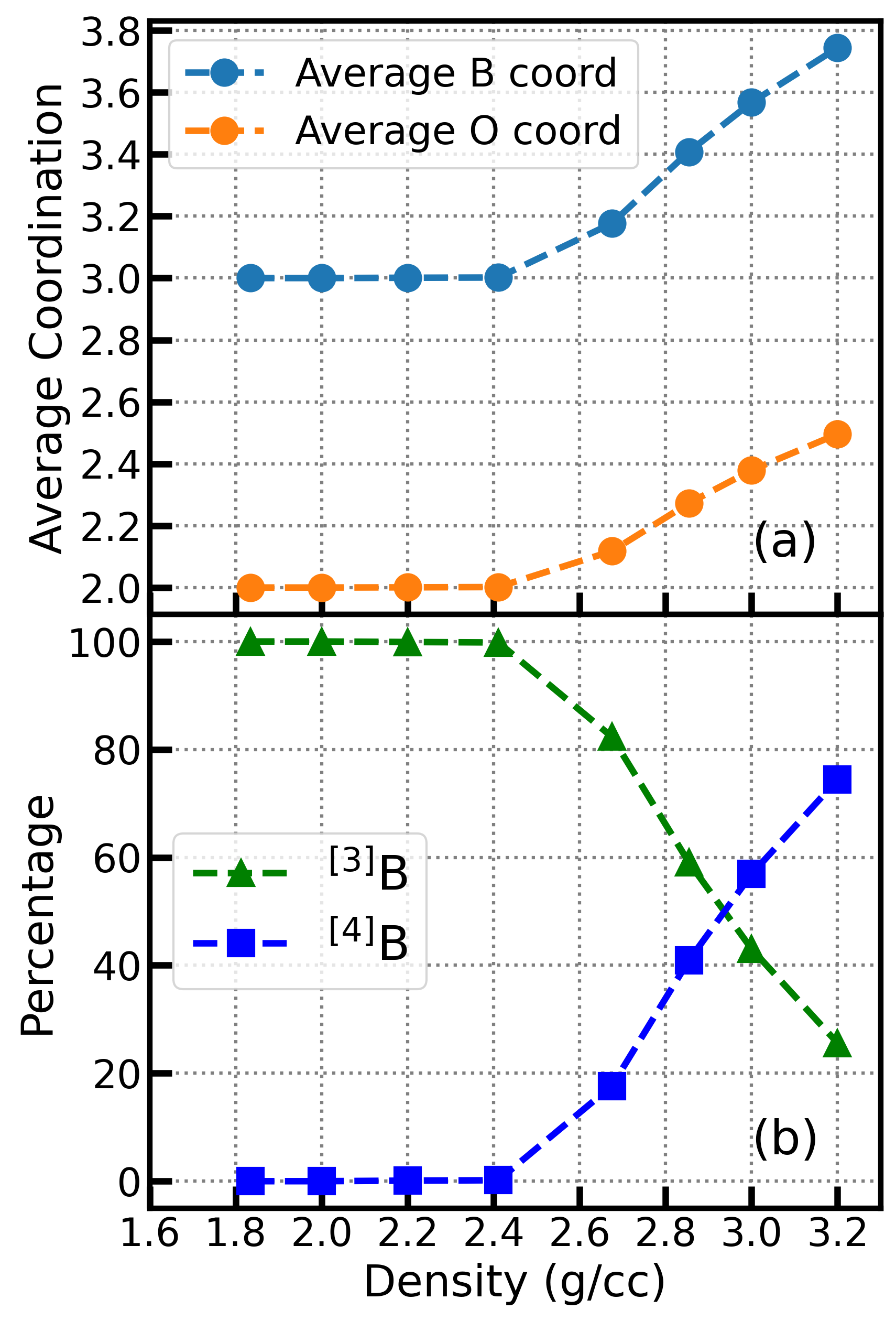}
   \includegraphics[trim= 0cm 9cm 6cm 0cm, clip, width=0.85\linewidth]{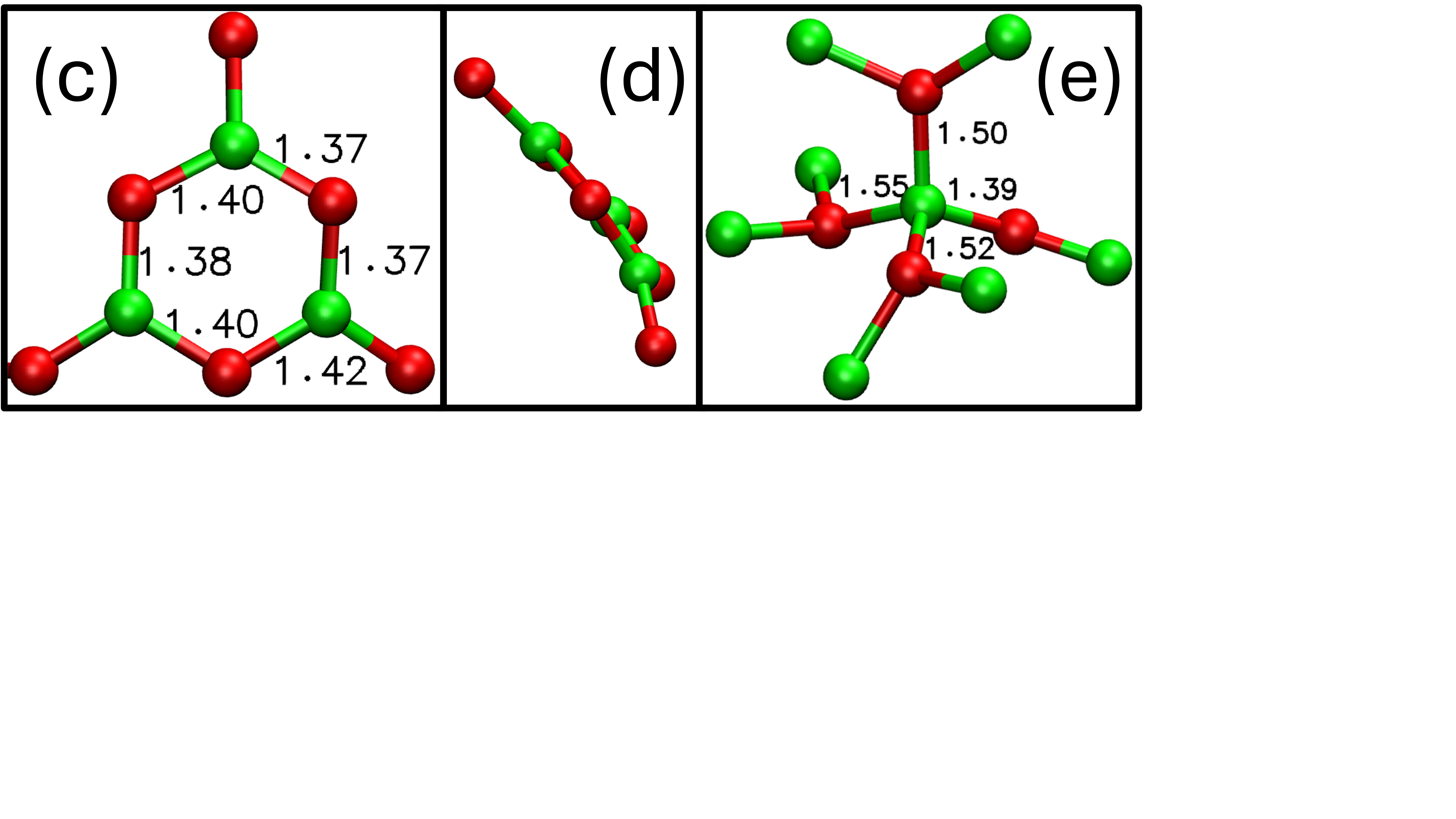}
    \caption{(a) Mean coordination number of boron and oxygen vs glass density. (b) Percentage of four- and three-coordinated boron atoms vs density. (c)-(d) Six-membered boroxol ring observed in the \bo\ glass network at 1.834 g/cc shown in two views. (e) Tetrahedrally coordinated boron was observed in the \bo\ glass network at 2.854 g/cc. Green: Boron, Red: Oxygen. Bond lengths in \AA\ are marked. The coordination numbers are obtained by averaging over four independent trajectories.}
    \label{fig:avg_coord_no}
\end{figure}
All borons are three-coordinated at ambient conditions, and all oxygens are two-coordinated. At 2.411 g/cc, the three-coordinated species constitute 99.9\% of the boron atoms. With increasing density, the fraction of four-coordinated boron increases (Figure~\ref{fig:avg_coord_no}(b)). At 2.854 g/cc, around 40\% borons are four-coordinated, and the fraction increases to around 74.4\% at a density of 3.2 g/cc. The same is reflected in Figure~\ref{fig:avg_coord_no}(a), wherein the mean oxygen coordination number increases beyond two at higher densities. 

\subsubsection{Angle Distributions}
\begin{figure}
    \centering
    \includegraphics[width=0.8\linewidth]{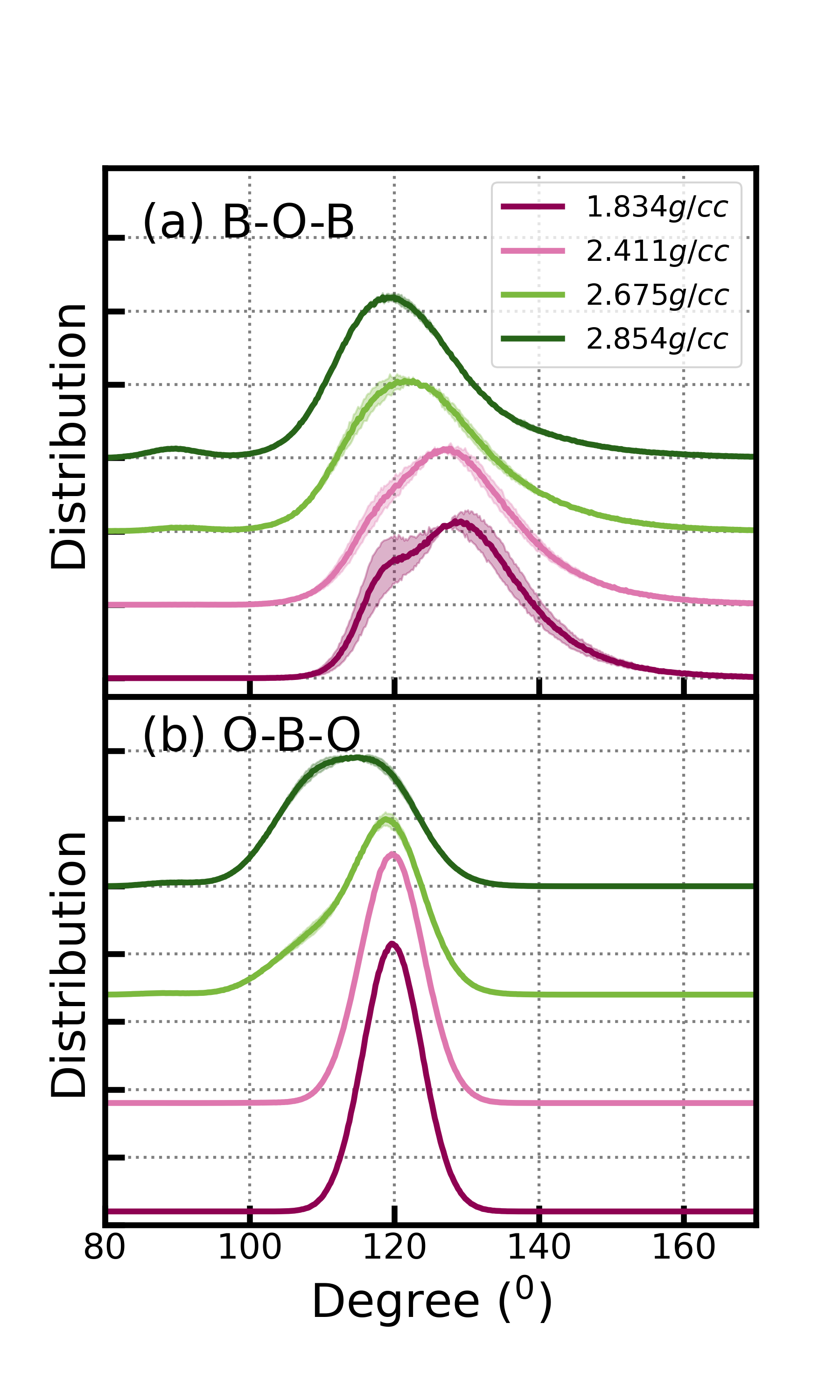}
    \caption{Bond angle distribution of \bo\ glass at 300 K formed with a quenching rate of 10$^{11}$ K/s. (a) B-O-B (b) O-B-O. The shaded region around lines shows the standard deviation over four independent trajectories.}
    \label{fig:ang_dist_press}
\end{figure}
The two bond angle distributions, B-O-B and O-B-O of glassy \bo\ at four different densities, provide insight into the local geometry of the key motifs in the network. The plotted angle in Figure~\ref{fig:ang_dist_press} represents the angle formed at the center of an atom by its nearest neighbors. As density increases, the distributions broaden, suggesting a larger variety of coordination geometries.
The distribution of B-O-B angles peaks at 135$^o$ and displays a prominent shoulder at 120$^o$ at ambient conditions. These have been attributed to out-of-ring and in-ring triplets, respectively~\cite{Ferlat_2008_Ring}. At 2.854 g/cc, this distribution changes to an unimodal one centered at 120$^o$, with a tiny peak at 90$^{o}$ whose intensity does not show a systematic dependence on the quenching rate (see Figure S8). Four-membered rings of four-coordinated boron and three-coordinated oxygen atoms contribute to the latter. Additionally, the peak in O-B-O angle distribution shifts from 120$^\circ$ to 109$^\circ$ with increasing density, which is a consequence of the formation of four-coordinated boron from a situation where all the borons were three-coordinated. 

\subsubsection{Analysis of Rings}
\begin{figure}
    \centering
    \includegraphics[width=1.1\linewidth]{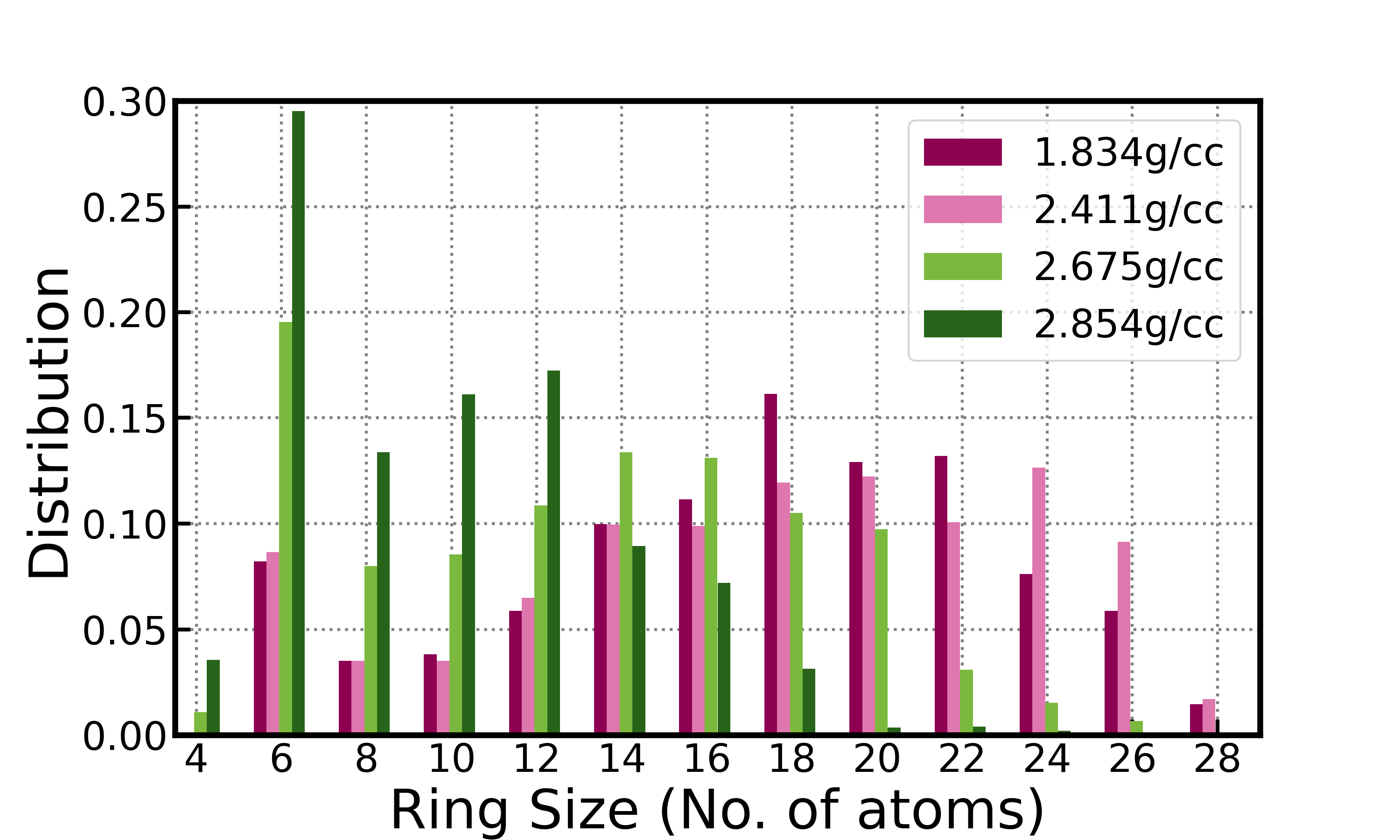}
    \caption{Ring size distribution in \bo\ glasses at different densities. With increasing density, the fraction of smaller rings increases.}
    \label{fig:ring_analysis}
\end{figure}

Figure~\ref{fig:ring_analysis} displays the distribution of ring sizes in \bo\ networks at the four densities. At 1.834 g/cc, there are a few six-membered rings, likely to be planar boroxol rings; we plan to pursue a detailed study of their occurrence and distinctive vibrational signatures~\cite{Ferlat_2008_Ring} in the future. The fraction of medium-sized rings (say 14-18) decreases with increasing density, while those of small rings increase, as expected. However, the dominant six-membered rings at 2.854 g/cc are not boroxol rings, which are constituted solely by three-coordinated boron atoms; at this density, a good fraction of borons are four-coordinated, thus forming a non-planar structural motif.

\section{Conclusions}

The present work has unraveled several features of \bo\ glass formation and structure at ambient and high pressures; much insight has also been obtained on methodological aspects of the modeling of glasses. These have been made possible through the systematic development of a machine learning potential derived from a collection of relevant atomic configurations spanning the range of thermodynamic conditions of interest, from 300 K-3000 K and 0 GPa-500 GPa and their total energies, forces on atoms and virial calculated using DFT at the revPBE-D3 level of theory. The MLP reproduced the smooth change in the coordination number of boron atoms from three at ambient conditions to four at high pressures corresponding to the formation of tetrahedral BO$_4$ motifs, consistent with experimental observations. The increase in boron coordination also led to a slight but non-negligible increase in the B-O bond distances from 1.37\AA\ in triangular BO$_3$ units to 1.45\AA\ in the four-coordinated units. Such subtle effects are hard to capture within a force field. Correspondingly, the angle distributions of O-B-O and B-O-B, too, changed with increasing pressure, reflecting the changes in the coordination geometries. 

It was easier for us to prepare pressure-quenched samples in the DPMD simulations than to quench the melt under constant pressure conditions. Thus, the simulations of \bo\ glasses at various densities ranging from 1.8 g/cc to 3.2 g/cc were performed under constant NVT conditions. The pressure-density relation from the simulations of \bo\ glass at 300 K, quenched at a conservative rate of 10$^{11}$ K/s, closely mirrors the experimental data. X-ray and neutron structure factors calculated from the model structures at various pressures too agree excellently with experimental results.

Earlier DFT-based simulations of \bo\ have used either the PW~\cite{pasquerello_prl_2005} or the PBE~\cite{JPCC_high_press_2024} functional, and the MLP developed to model borosilicate glass~\cite{Urata_LBS_2022} employed the PBE functional. Ferlat et al. showed the importance of including van der Waals corrections to the PBE functional to correctly predict the density and stabilization energies of crystal polymorphs of \bo\ \cite{ferlat2019van}. The MLP developed here is based on the revPBE functional with Grimme's D3 van der Waals corrections~\cite{dispersion-D3}. While this level of theory has been widely used for the study of molecular liquids, its suitability for an extended network solid such as \bo\ needs to be checked against reference calculations on clusters using higher levels of theory~\cite{li2013first}. 

The current work has also highlighted subtle but often-ignored aspects of force field and/or explicit DFT-based MD simulations of glasses. The latter are typically run for several tens of picoseconds, which may not be sufficient for the evolution of the structure from that used in generating the initial configuration -- such as a force field. The result presented for the B-O-B angle distribution in Figure~\ref{fig:AIMD_comparison} brings to light this fact. On the other hand, force fields can potentially underestimate the diffusion coefficient of species even in the melt. This characteristic is shared by several force fields that do not incorporate polarizability explicitly, whose manifestation as reduced diffusion of species is particularly evident in ionic fluids~\cite{avula2021efficient}. The \bo\ melt modeled with MLP-26 shows facile diffusion of ions; in contrast, under the same conditions, ions modeled with the force field are barely mobile (see Figure S7). 

Glassy networks modeled with AIMD can also suffer from system size limitations which can potentially reduce the dimensions of the rings in the structure. Herein, we carried out a limited investigation of the same by performing DPMD simulations using two system sizes -- 1700 atoms and 360 atoms. The distribution of ring sizes in the latter (Figure S11) reduced to zero abruptly, while that in the former did so smoothly (Figure \ref{fig:ring_analysis}), demonstrating the need for system sizes adequate to accommodate the largest ring dimensions. 
Analysis of the glassy \bo\ network at 1.834 g/cc revealed rings of several sizes whose statistics change with pressure. The planar, six-membered boroxol ring is widely believed to dominate at ambient conditions; these, too, were observed in the DPMD simulated structures. The exact proportion of such and its vibrational (Raman scattering) signature will be examined in detail. As has been alluded to in Ref.~\citenum{Ferlat_2008_Ring}, the prevalence of the boroxol ring could be dependent on the quenching rate, even if the underlying potential energy surface is accurate (such as in the present instant). DPMD based on MLP makes such an interesting study of the dependence of their fraction on the quenching rate possible.

A crucial result of the current work is found in the dependence of the density of the glass on the quenching rate, particularly at high pressure. At a given pressure, we observed increased densification of the glass with a decreasing quenching rate, an observation consistent with experiment~\cite{SOPPE1988201}. All the results of the structure of \bo\ presented here are obtained at a conservative rate of 10$^{11}$ K/s, one which is just not possible to be adopted in an AIMD simulation. We plan to pursue investigations at even slower quenching, particularly to study intermediate-range order. Quenches at rates higher than 10$^{11}$ K/s are shown herein to lead to artifacts in the structure.

The current work thus paves the way for a slew of further investigations into the diversity displayed by glassy \bo\ in its coordination environments, intermediate range structure, and their impact on vibrational modes and ion transport. These will constitute the objectives of our future work. 

\section*{Supplementary Material}
More details about the development of the machine learning potentials, comparisons with AIMD, system size dependence, and example input files are given in the supplementary material.

\section*{Acknowledgements}
This work was partly funded by a grant from the Board of Research in Nuclear Sciences (BRNS), India (Grant No.: 58/14/06/2020-BRNS/37060).  DM thanks UGC, India for a fellowship. The support and the resources provided by the 'PARAM Yukti Facility' under the National Supercomputing Mission, Government of India, at the Jawaharlal Nehru Centre For Advanced Scientific Research are gratefully acknowledged.

\section*{Author Declarations}
\subsection*{Conflict of Interest}
The authors have no conflicts to disclose.

\subsection*{Author Contributions}
{\bf SB:} Conceptualization (equal); Funding Acquisition (lead); Methodology (equal); Supervision (lead); Writing (equal).
{\bf NVSA:} Conceptualization (equal);  Formal Analysis (supporting); Methodology (equal); Investigation (supporting); Writing (equal).
{\bf DM:} Data Curation (lead); Formal Analysis (lead); Methodology (equal); Investigation (lead);  Visualization (lead); Writing (equal).

\section*{Data Availability}
The data that support the findings of this study are available within the article and its supplementary material.

\section*{References}
\bibliographystyle{aipnum4-1}  
\bibliography{ref}

\end{document}